# On the features of the Optical Rogue Waves observed in the Kerr lens mode locked Ti:Sapphire laser.

Alejandro A. Hnilo, Marcelo G. Kovalsky and Jorge R. Tredicce (*).

*Centro de Investigaciones en Láseres y Aplicaciones (CEILAP), CITEDEF, J.B. de La Salle 4397, (1603) Villa Martelli, Argentina.*
*(*) Pole Pluridisciplinaire de Materiaux et Energie (PPME), Université de la Nouvelle Caledonie, Nouméa, Nouvelle Caledonie.*
*emails*: ahnilo@citedef.gob.ar, alex.hnilo@gmail.com

### *Abstract*

Kerr lens-mode-locked Ti:Sapphire lasers are known to display three coexistent modes of operation, that can be described as: continuous wave (CW), transform limited pulses (P1) and positive chirped pulses (P2). Optical rogue waves, in the form of pulses of high energy appearing much often than expected in a Gaussian distribution, are observed in the chaotic regime of the mode P2, but not of P1. These high energy pulses appear in an unpredictable way, but it is observed that their separation (if measured in number of round trips) can take only some definite values, which received the name of "magic numbers". The existence of optical rogue waves in P2 and not in P1, and also of the magic numbers, are correctly reproduced by a numerical simulation based on a five-variables iterative map. But, a successful numerical simulation provides limited insight on the physical causes of the observed phenomena. We present evidence that optical rogue waves in this laser follow a modulational instability, and that an initial condition in P1 rapidly evolves into P2 if the parameters' values are beyond that instability's threshold. The magic numbers are residuals of the periodic orbits of the "cold" cavity when it is perturbed by the opposite effects of a dissipative term, due to the presence of transversal apertures, and an expansive term, due to the Kerr effect.

March 20th, 2014

# 1. Introduction.

Waves of extremely high amplitude, appearing outside the Gaussian distribution, are important phenomena in deep ocean waters, and received the name of "freak", or "rogue", waves [1]. In the last decade, scientific interest increased on analogous rare events of large amplitude observed in areas other than Oceanography. In this sense, "Optical rogue waves" (ORW) were first observed in the light intensity fluctuations at the edge of the spectrum produced by ultrashort pulse pumped, micro-structured optical fibers, in the threshold of super-continuum generation [2,3]. Conditions for their formation were determined in specially designed experiments using optical fibers [4]. ORW were observed in a VCSEL with an injected signal [5], in all-solid state lasers with a (slow) saturable absorber [6] and in the Kerr-lens mode locked (KLM) Ti:Sapphire laser [7] (fast saturable absorber). It is worth remembering here that the usual approach for the "laser + absorber with a nonlinear response" problem involves equations of the Nonlinear-Schrödinger type (NLSE), which is also usual in the description of rogue waves in the seas. Therefore, it is not surprising that similar extreme-event phenomena are observed in both systems.

The quantitative definition of an extreme event usually is: *a)* amplitude higher than twice the average calculated among the set of the 1/3 highest events in the series (the "abnormality index", AI); or, *b)* amplitude higher than four (sometimes, eight) times the standard deviation. These two definitions can be coincident or not, depending on the form of the distribution. They are precise, but quite arbitrary. The optical community has often employed the additional criterion of a long tailed or L-shaped distribution. In previous contributions, we have also used the value of the kurtosis as an additional quantitative measure of a non-Gaussian feature (a Gaussian distribution has kurtosis = 3).

The KLM Ti:Sapphire laser is the most widely used source of ultrashort (femtosecond, fs) light pulses nowadays. The dynamics is intrinsically complex, because it is ruled by a delicate balance of several spatial and temporal effects. In the temporal domain the group velocity dispersion (GVD) in all the optical components, and the intensity dependent self-phase modulation (SPM), mostly in the laser rod, are balanced by the dispersion produced by an intracavity pair of prisms. In the spatial domain, the relevant features are the cavity's geometrical configuration, the beam diffraction, and the intensity dependent self focusing (SF). The amplification in the active medium is an additional source of nonlinearity through gain saturation. The pulse intensity, the beam size and the pulse duration are coupled by both the SPM and the SF. Three coexistent dynamical modes of operation are observed in the laser output: continuous wave (CW), transform limited pulses (P1), and positive chirped pulses (P2). The laser spontaneously evolves from one pulsed mode to the other even in the absence of noise [8]. In the lab, one can induce a transition by mechanical perturbations. As the GVD of the laser cavity is adjusted close to zero (coming from the negative side), the pulsed modes evolve towards chaos following a different route: P1 through quasiperiodicity, P2 through intermittency [9]. Be aware that what we call "mode" here means a *dynamical* mode of operation, not a spatial cavity mode (as, say, a Gauss-Laguerre mode).

Regarding the ORW observed in this laser, two phenomena appear especially intriguing:

*i)* ORW are observed in the chaotic regime of P2, but not of P1.

*ii)* Even though the appearance of an ORW seems to be unpredictable, preferred numbers are observed in the distance between successive ORW (as measured in cavity round trips, or intermediate non-ORW pulses). We named them "magic numbers" [10].

Both phenomena are accurately reproduced by a numerical simulation based on a five-variables iterative map. This map relates the values of the beam size and curvature, pulse duration, chirp and energy, of the *n+1* pulse in the mode locking train, to the values of the same variables in the previous *n*-pulse. Yet, a numerical simulation, even if it is successful, provides limited intuitive insight on the physical causes. In this paper, we present evidence that ORW in this laser appear after crossing the threshold for the modulational instability, what makes this phenomenon even closer to the rogue waves in the ocean. We find that ORW are not observed in the P1 mode because, at the parameters' values beyond that threshold, this mode is unstable and the system rapidly evolves into the (more stable) P2 mode. The magic numbers, on its turn, are found to be the residuals of the periodic orbits of the cold laser cavity (i.e., the optical cavity without gain, losses, SF or SPM) when perturbed by opposite dynamical effects. A simplified map in the complex plane predicts that this occurs when the losses due to transversal apertures and the nonlinear Kerr effect hold to a particular condition, which is precisely fulfilled in this laser.

The study of ORW in KLM lasers is a subject of dynamical systems which is interesting by itself. Besides, it will lead to a deeper understanding of the operation of these lasers and, eventually, to an improvement of their performance. As the analogies with the rogue waves in the ocean appear more definite, KLM lasers may provide a convenient test bench to study the general features of the essentials of the phenomenon of extreme events. For, lasers evolve much faster and are much easier to observe than the wave dynamics in the open seas. On the other hand, one of the motivations to study "real" rogue waves is the potential damage they can inflict to ships or platforms. In the best of our knowledge, there are no reports on damages produced in optical systems by ORW in Ti:Sapphire lasers. In this sense, it is fortunate that ORW exist only in the P2 mode, because the P1 mode is usually the preferred one in the practice. The fact that Ti:Sapphire lasers are seldom operated in the P2 mode may be then the reason why catastrophic consequences of ORW have not been reported.

In the next Section, we describe the experimental setup where ORW are observed, and review the theoretical description with the five variables iterative map. In the following Section 3, we discuss the modulational instability for modes P1 and P2. In the Section 4, we review the problem of the magic numbers and propose and test a solution in terms of the orbits of a simplified map in the complex plane.

**2. Background.**

*2.1 The experiment: setup and main results.*

The scheme of our Ti:Sapphire laser is shown in the Fig.1. It is a standard "X" configuration, with a flat HR rear mirror (M4) and a 12% output coupler (M1). The total cavity length is 1724mm (cavity frequency: 87 MHz). For 5W CW pump at 532nm, (provided by an intracavity frequency doubled, diode pumped Nd:YVO$_4$ laser) the typical output power is 400 mW in the spectral region around 800nm. Typical pulse durations for uniform mode locking are 35 fs (mode P1) and 65 fs (mode P2), as measured with an autocorrelator. We observe the mode-locking pulse train with a fast photodiode (0.2mm diameter, 0.5 ns risetime) and record the signal in the memory of a high speed sampling oscilloscope. The photodiode is too slow to resolve the fs pulse shape, what we observe is the instrumental response to a "delta" input. The height of the observed electric pulse is proportional to the total energy of the laser pulse. Yet, there are about only 23 samples in a round trip time, and most of them just draw the base line between successive pulses and are useless to study the dynamics. Only a few samples are available to draw each electric pulse and, of course, they rarely coincide with its real peak. In order to find the real value of the peak of each pulse, we reconstruct its shape using a fourth-order polynomial algorithm to find the best fitting to the ten samples closest to the highest one. Then, the value of the maximum of the fitting curve is taken as the real value of the peak, for each pulse in the train. This method provides time series with noise low enough to allow the calculation of the dimension of embedding of the attractors and the values of the Lyapunov exponents, and even an estimation of the fractal dimension in favorable regions of the parameters' space.

As said before, this laser shows two coexisting self-pulsing modes of operation, P1 and P2. In the bistable region of the parameters, the laser wanders from one to the other mode in a time scale of several minutes. Special care is taken here that the laser operates always in a single transverse, near TEM$_{00}$ cavity mode. Otherwise, one may be confused by similar in appearance, but completely different spatio-temporal effects. The dynamical modes P1 and P2 can be distinguished by the pulse duration, the chirp, the spectrum and, even by the naked eye, as a change in the size of the laser spot. The pulses of the mode P1 are transform-limited, and twice shorter than the P2 ones. The average SPM is larger for P2 because the pulse for this mode is shortest at the center of the laser rod (hence producing larger SPM in the propagation inside the rod), while for P1 the pulse is shortest at the output end of the laser rod. Besides, the pulses in the P2 mode carry more energy in the average.

If the GVD is adjusted close to zero (coming from the negative side), each mode follows its own road to chaos: P1 through quasiperiodicity, P2 through intermittency [9]. Once in the chaotic regime of P2, ORW are easily observed (Fig.2), for all the usual criterions defining an extreme event described in the Introduction [7]. Typically, 100 to 200 pulses from $10^4$ qualify as ORW. On the other hand, no ORW are observed in the mode P1, for any accessible value of the GVD. This result supports the hypothesis of a non-trivial and deterministic nature of the observed ORW. For, if the ORW in P2 were mere noise, or caused by self-Q-switching, there is no reason why they would not be observed in the coexisting mode P1 too. The theoretical approach based on the numerical running of a five variables iterative map, which is described in the next section, predicts the existence of ORW for P2 and not for P1, hence agreeing with the observations.

The hypothesis of a deterministic origin for these ORW is strengthened by the observation that, even though their emergence is apparently unpredictable, they tend to appear at definite distances of each other. The Fig.3 shows the histograms of the distances between successive ORW, measured as the number of the laser cavity round trips, for a representative time series. It is immediately seen that they are not uniformly distributed, but that they take only some preferred or "magic" numbers: {11, 12, 23, 35, 46, 58, 94} in the experimentally obtained histogram, and {11, 12, 23, 24, 34, 35, 46, 58} in the theoretically obtained (from the five-variables iterative map) one. Note the remarkable coincidence between the experimental and the theoretical distributions. Then, it is possible to predict with a high probability the moment in which an ORW will not be produced, despite the dynamics is chaotic. Note also that the magic numbers are simple combinations of the numbers 11 and 12. There are, however, "missing" combinations, as 22, 57 and 69 in both sets. Besides, there is an internal structure: if the distance between the *n*-ORW and the *(n+1)*-ORW is 11, there is a 93% probability that the distance to the *(n+2)*-ORW is 12 (this probability is 77% in the theoretical series). In the same way, 23 follows 12 (with probability 82% both in the experimental and the theoretical series), 12 follows 23 (66% experimental and 40% theoretical), 23 follows 35 (61% experimental and 56% theoretical), and after 58 comes 35 in 100% of the cases (both experimental and theoretical). Second-step *n* to *n+2* correlations also exist, but they are weaker. Perhaps contrary to intuition, there is no correlation between the amplitude of an ORW and its magic number [10]. The numerical coincidences between the experimental and the theoretical results indicate that the magic numbers have a deep and robust cause, and that noise plays a minor role in the dynamics of ORW in this type of lasers. It is to be noted that no fine tuning of the many laser parameters has been performed in the numerical simulations. Tabulated and directly measured values have been used. Only the GVD has been adjusted, to fit the observed average pulse duration.

*2.2. The description with five-variables maps.*

The usual approach to KLM is through some version of the master equation for passive mode locking [11]. This is an equation for the slowly varying envelope of the electric field in partial derivatives with an intensity-dependent nonlinear term, of the NLSE type. Partial or full numerical approaches draw the pulse shape and provide the stationary values of the pulse variables, but the stability of the solutions is difficult to obtain. A standard approach to a nonlinear system is to reduce the continuous time dynamics to an associated discrete system: the iterative, stroboscopic or Poincaré map. The description with maps is alternative to that with a differential equation, and no information is gained or lost. There are, however, some immediate advantages: the dimensionality of the problem is reduced in (at least) one and the numerical simulations run easier and faster. However, writing the map equation can be as difficult as solving the differential equation, unless the physical system has some "internal clock" that determines the position of the adequate discrete times. In the case of passive mode-locked lasers, that clock is provided by the cavity round trip time and, in fact, it is simple to obtain recursive equations linking the pulse variables in the *n+1*-round trip with the values taken at the *n*-round trip [12]. The stability of periodical solutions (as the uniform mode locking pulse train) is easily determined. There are additional advantages when studying unstable behaviors, for there is no theoretical limitation on the acceptable pulse variation from one round trip to the next, and the observed period-doubling solutions are trivially described with maps.

In order to derive the description of KLM lasers with maps, we suppose that the electric field inside the mode locking pulse is given by: $E(t) = E_0 \exp(-i\ kr^2/2q) \exp(-i\ kt^2/2p)$ where *r* is the distance to the optical axis, *k* is the wavenumber and *p, q* are the beam parameters defined by the relationships $1/q = n/R - i2k/\sigma^2$ and $1/p = Q - i2k/\tau^2$, where *n* is the index of refraction of the medium, and the pulse variables are: $\sigma$ the spot radius, *R* the beam curvature radius, $\tau$ the pulse duration and *Q* the chirp. As the pulse propagates through an optical element or distance, *q* changes according to:

$$q_{out} = (\mathbf{A}.q_{in} + \mathbf{B})/(\mathbf{C}.q_{in} + \mathbf{D}) \qquad (1)$$

where {**A**...**D**} are the elements of a 2x2 matrix. An analogous relationship holds for *p* [13]. The total round-trip matrix describing the effect of the many elements inside a laser cavity is obtained by multiplying the matrices of each element. In general, the propagation is fully described by 4x4 matrices. The Gaussian approximation implicit in this approach is generally valid for pulses longer than 10 fs. Under appropriate design conditions, which usually hold for Ti:Sapphire laser cavities, the general 4x4 round trip matrix can be split into two 2×2 diagonal blocks, which we call "spatial" matrix (**ABCD**) and "time" matrix (**KIJL**):

$$M = \begin{pmatrix} A & B & 0 & 0 \\ C & D & 0 & 0 \\ 0 & 0 & K & I \\ 0 & 0 & J & L \end{pmatrix} \quad (2)$$

The matrix elements include terms due to the SF and SPM in the laser rod, which are, in turn, functions of the already mentioned four pulse variables and the pulse energy $U$. These terms, named here *nonlinearities*, have the form $\gamma = c_\gamma \cdot U/(\tau\sigma^4)$ for the spatial matrix and $\beta = c_\beta \cdot U/(\tau^3\sigma^2)$ for the time matrix. At first order in the nonlinearities, both sub-matrices have determinant equal to 1. The constants $c_\gamma$ and $c_\beta$ are proportional to the nonlinear index of refraction of the Ti:Sapphire and their precise expression is rather involved [14]. Their numerical values for our laser are $c_\gamma = 1.38\times10^{-11}$ cm$^4$ fs/nJ and $c_\beta = 2.18\times10^{-7}$ cm$^2$ fs/nJ.

It is convenient defining new variables: $S\equiv 1/\sigma^2$, $T\equiv 1/\tau^2$, $\rho\equiv 1/R$. The expressions that link the pulse variable values at the *n+1* round trip with the ones at the *n* round trip are then:

$$S_{n+1} = \frac{S_n}{(\mathbf{A}+\mathbf{B}\rho_n)^2 + (\mathbf{B}\lambda S_n)^2} \quad (3a)$$

$$\rho_{n+1} = \frac{(\mathbf{A}+\mathbf{B}\rho_n)(\mathbf{C}+\mathbf{D}\rho_n) + \mathbf{BD}\,(\lambda S_n)^2}{(\mathbf{A}+\mathbf{B}\rho_n)^2 + (\mathbf{B}\lambda S_n)^2} \quad (3b)$$

$$T_{n+1} = \frac{T_n}{(\mathbf{K}+\mathbf{I}Q_n)^2 + (\frac{\mathbf{I}T_n}{\pi})^2} = T_n \frac{\mathbf{L}-\mathbf{I}Q_{n+1}}{\mathbf{K}+\mathbf{I}Q_n} \quad (3c)$$

$$Q_{n+1} = \frac{(\mathbf{K}+\mathbf{I}Q_n)(\mathbf{J}+\mathbf{L}Q_n) + \mathbf{IL}(\frac{T_n}{\pi})^2}{(\mathbf{K}+\mathbf{I}Q_n)^2 + (\frac{\mathbf{I}T_n}{\pi})^2} \quad (3d)$$

$$U_{n+1} = U_n \left\{ 1 - \frac{2(U^* S_n + U_n S^*)}{\mu D_s} + 4(\mu - 1)/\mu \right\} \tag{3e}$$

where $\mu$ is the product of the small signal gain and the single passage feedback factor due to passive losses ($\mu$=1.61 for the working point of this laser), and $D_s = 1.22$ mJ/cm$^2$ is the saturation energy flux (i.e., the saturation energy multiplied by the cavity round trip) for Ti:Sapphire.

The matrix elements in (3) are expressed as a series expansion on the nonlinearities:

$$\mathbf{A} = \mathbf{A}_0 + \gamma\, \mathbf{A}_\gamma + \gamma'\, \mathbf{A'}_\gamma + \gamma^2\, \mathbf{A}^{(2)}_\gamma + \gamma'^2\, \mathbf{A'}^{(2)}_\gamma + \gamma\gamma'\, \mathbf{A''}^{(2)}_\gamma + ... \tag{4a}$$

the same for **B**, **C**, **D**, and:

$$\mathbf{K} = 1 + 2\delta\beta' \tag{4b}$$
$$\mathbf{I} = 2\delta \tag{4c}$$
$$\mathbf{J} = 2\delta\beta\beta' + \beta + \beta' \tag{4d}$$
$$\mathbf{L} = 1 + 2\delta\beta \tag{4e}$$

where $2\delta$ is the value (negative) of the GVD per round trip. The coefficients of the expansions (4a) are functions of the geometrical parameters only. The factors $\gamma$ ($\gamma'$) and $\beta$ ($\beta'$) are the nonlinearities when the pulse crosses the rod from $M_3$ to $M_2$ (from $M_2$ to $M_3$), see the Fig.1.

The recursive relations (3) with the matrix elements given by (4) define an iterative map that correctly describes much of this laser's dynamics. The fixed points can be obtained analytically at first order in the nonlinearities. The fixed point $T_n = T_{n+1} = 0$ ($\tau \to \infty$) makes the nonlinearities zero and corresponds to the CW mode. The fixed point $Q_n = Q_{n+1} = 0$ leads to $\beta = \beta'$ and corresponds to the P1 mode. There is a fixed point with positive chirp, for which $\beta \gg \beta'$ (P2 mode) and one with negative chirp, for which $\beta \ll \beta'$ (P3 mode). The stability analysis of the fixed points explains why the P3 mode is never observed in the practice. The P2 mode has higher energy and longer pulses than P1, and its region of stability is larger. The slope of pulse duration vs. GVD (which is nearly a straight line for the usual GVD operation values) is twice larger for P2 than for P1.

For reasons of speed and simplicity of the numerical simulations, we use three different iterative maps. In one of them the condition $\beta = \beta'$ is enforced, and hence it describes the evolution of the P1 mode only, as if the P2 mode didn't existed. We call it the "P1-map". In the same way, the forced condition $\beta' = 0$ defines the "P2-map". A more involved numerical simulation, where $\beta$ and β' evolve freely, is the "Bistable-map".

## 3. Optical rogue waves in the different modes.

*3.1 The modulational instability.*

It is a description of instabilities in ocean waves that is believed to be a possible mechanism for the formation of rogue waves [15,16]. The starting point is the NLSE:

$$\partial A/\partial z + ig\partial^2 A/\partial t^2 - i\eta|A|^2 A = 0 \qquad (5)$$

where $A$ is the envelope wave amplitude, $g$ is the GVD per unit length and $\eta$ is the Kerr nonlinear coefficient per unit power and length. A periodical solution is assumed for $A(t)$, and a harmonic perturbation of frequency difference ω with the periodic solution is introduced. The linear stability analysis of this perturbation shows that an eigenvalue becomes positive (i.e., the perturbation grows exponentially) if $g<0$ and:

$$\eta|A|^2 > \tfrac{1}{2}|g|\omega^2 \qquad (6)$$

or, in other words, that the (focusing) Kerr nonlinearity must overwhelm the spread produced by the GVD (scaled with the frequency of the perturbation).

The self-mode-locked laser is also usually described by a NLSE-type equation [11], where a short-time variable describing the pulse shape plays the role of *t* in (5), and a long-term variable describing the pulse-to-pulse modulation plays the role of *z*. For large amplifier's bandwidth, which is valid for the pulse durations of interest here (>10 fs), and neglecting the phase and frequency shift per round trip and after the level of gain saturation is reached, which are all valid approximations in the uniform mode-locking regime, the equation for the self-mode-locked laser becomes formally identical to (5). The amplitude *A* is now the slowly varying envelope of the electric field, the periodic solution of *A* is the uniform mode-locking signal, and the condition (6) is that the SPM contribution per round trip must be larger than the scaled GVD per round trip, $\beta|2\delta| > \tfrac{1}{2}|2\delta|\omega^2$ or:

$$\beta = c_\beta \cdot U/(\tau^3 \sigma^2) > \tfrac{1}{2}\omega^2 \qquad (7)$$

As β is, in the average, larger in the mode P2 than in the P1, it is conceivable that P2 is able to cross the modulational instability threshold (and then to evolve into an ORW regime) while P1 is not. In support of this idea, we note that the modulational instability is known to generate spectral sidebands, what coincides with the observed spectral shape of the mode P2.

To explore this tentative explanation of the formation of ORW in this laser, the values of β and U for each pulse in a train of $10^5$ are plotted in the Fig.4, for each of the two modes. The P1(P2)-map is used to calculate the points corresponding to the P1(P2) mode. The ORW (mode P2) are the dots at the right of the vertical dotted line. Note that, even though the P2 pulses have a large dispersion, all the ORW have practically the same value of β (what we call here $β_{ORW}$) which is, curiously, relatively low for the mode P2 ($β_{ORW} ≈ 10^{-6}$ fs$^{-2}$). The pulses of the mode P1, instead, accumulate in a small "cloud", just below that value. It is interesting the spiraling orbit corresponding to a quasiperiodical excursion, typical of the P1 mode. Excepting for a few pulses of very low energy (on the left), the whole population of the P2 mode has a larger value of β than the highest of the P1 mode.

As it is seen, some pulses of the P1 mode are just at the border of reaching the apparent threshold of the modulational instability, which we speculate it is given by the relatively well defined value $β_{ORW}$. Apparently, a small increase in the pulse energy may allow the P1 mode to reach the ORW regime. We then increase the small signal gain in the P1-map, and plot histograms of the energy pulse distribution. A 10% increase (Fig.5b) changes the shape of the distribution (compare with Fig.5a): a central maximum appears and a high energy tail starts to develop. For a 20% increase the tail is longer (Fig.5c) and, for a 40% increase, ORW are obtained. As a further check, the values of β,U are plotted in the Fig.6. In comparison with the Fig.4, the form of the P1-cloud has changed and most of the pulses are above the presumed threshold value $β_{ORW}$, as it had been the case for the P2 with the unmodified value of the small signal gain.

*3.2 Stability of the P1 mode in the ORW regime.*

At this point, one is tempted to conclude that ORW should be observed also in the P1 mode if the small signal gain (in the practice: the pump power) were sufficiently increased. This conclusion may have important practical consequences for, as it is said in the Introduction, P1 is the mode of operation preferred by this laser's users, and an ORW in a high power fs laser may have catastrophic consequences in any optical setup. Nevertheless, we suspect that the conclusion may be erroneous. We were not able to increase the total pump power in the practice, but we did study the laser's dynamical behavior under many different pump focusing and alignment conditions. In no case we observed ORW in the mode P1, and in almost all cases we observed them easily in the mode P2. We think implausible that the instability threshold was never crossed for P1, for the gain conditions of modulational instability for P1 and P2 are, after all, not so far away from each other.

Roughly speaking, a 40% increase in the small signal gain means a 20% reduction in the pump focusing radius, a condition that we should have reached, even if inadvertently, in some occasion. On the other hand, the study of the volumes of the basins of attraction and of the change of the eigenvalues, as the laser parameters are varied, indicates that the P1 mode is generally less stable than the P2 [9]. This is confirmed by the operation of this laser in the practice.

Therefore, as an additional check, we run numerical simulations with the bistable-map, with initial conditions in the ORW regime of the P1 mode, and follow the pulse evolution (recall that the previous simulations were run with the P1-map). An example is shown in the Fig.7, starting with an initial condition in the fixed point of the P1 mode. The pulse values of the P1 mode are the cloud on the left, and the ones of the P2 mode are the cloud on the right, note the different average values of the variables of each cloud. After a few hundred iterations, corresponding to some μs of real time, the system crosses from the P1 initial condition to the P2 region and remains there. In no case the system remains in the P1 mode for a time long enough to be observed in the practice. We are therefore prone to conclude that ORW are not observed in the mode P1 because this mode is unstable, against the coexisting mode P2, if the laser parameters' values are adjusted above the modulational instability threshold. Therefore, we predict that no increase of the pump power can produce observable ORW in the P1 mode.

## 4. On the cause of the magic numbers.

*4.1 Changes if the threshold of ORW is lowered.*

The formation of an ORW in this laser is apparently unpredictable but, once an ORW emerges, it does not do it at an arbitrary distance from the previous ORW. As it was anticipated in the Section 2.1, the distance between successive ORW measured in number of round trips (or in numbers of intermediate non-ORW pulses) is always a simple combination of the numbers 11 and 12. On the other hand, as it was mentioned in the Introduction, the boundary between an ORW and a non-ORW is somewhat arbitrary. If this boundary is lowered the number of ORW increases, but the "new" ORW still appear at a distance that is a combination of 11 and 12. Noteworthy, these new ORW appear intermediate of the old ones, so that the higher magic numbers depopulate as the boundary is lowered.

F.ex., an experimental time series generates the histogram of ORW of the Fig.3a if the boundary is established, in accordance with the 2×AI criterion, as 87 arbitrary units (a.u.). The total number of ORW is 237 and the magic numbers are {11, 12, 23, 35, 46, 58, 94}. If the boundary is lowered to 80 a.u., the number of ORW increases to 328 but the highest magic numbers disappear, for the set is now {11, 12, 23, 24, 35}. If the boundary is further lowered to 75 a.u., the number of ORW increases to 397, the magic numbers 24 and 35 disappear and the number "1" appears in the set instead (i.e, there are two successive ORW). The same phenomenon of disappearance of the higher magic numbers occurs in the theoretical time series, and with a remarkable numerical coincidence with the observed ones. It seems clear that some robust quasi-periodical phenomenon is underlying, with a typical period related with 11 and 12 round trips.

*4.2 Periodicities of the spatial part of the map.*

The evolution of the spatial part of the pulse (i.e., its beam size and curvature radius) in absence of dynamical effects is a periodical phenomenon, and it is the first candidate to explain the origin of the magic numbers. In order to study this possibility, it is convenient to use a simplified form of the iterative map described before. The **ABCD** matrix of the Section 2.2 can always be transformed in such a way that **A**=**D**, so that the matrix has only two independent parameters, $A$ (with no units) and $B$ (unit of length). Using $B$ to scale $q$, such that $\psi \equiv B/q$, the eq.(1) becomes [17]:

$$\psi_{n+1} = A - (A + \psi_n)^{-1} \tag{8}$$

This map describes the evolution of the spatial parameter in the absence of gain, losses and nonlinear effects, what is called the "cold" cavity. The fixed points of (8) are $\pm i(1-A^2)^{1/2}$, one is physically meaningful and the other one is not. Fixed points of higher periodicity are found from the map after $n$ iterations starting from an arbitrary initial condition $\psi_0$:

$$\psi_n = (P_1^n + P_2^n \psi_0) / (P_2^n + P_3^n \psi_0) \tag{9}$$

where the $P_j^n$ are polynomials in $A$, similar to Legendre polynomials, whose form is detailed in [17]. The recursive relationship to calculate the $P_3^n$ is:

$$P_3^{n+1} = 2A\, P_3^n - P_3^{n-1} \tag{10}$$

The first few $P_3^n$-polynomials and their roots until $n = 13$ are displayed in the Appendix. If $P_3^n=0$ then $P_1^n=0$, and from (9) the solution is $n$-periodic for all initial conditions. It is then sufficient finding the zeroes of $P_3^n$ to know the periodicities of the cold cavity.

As is well known, the two fixed points of the map (8) have indifferent stability. By taking into account that transversal apertures unavoidably exist (due to the finite diameter of the mirrors, of the pumped region, etc.), the physically meaningful fixed point becomes definitely stable, and the unphysical one, unstable [18]. The effect of an aperture of radius $R_{ap}$ is taken into account in (8) by adding a term $-ia$, where $a=2B/kR_{ap}^2$. With this modification, the map (8) converges to the physically meaningful fixed point (whose expression is different now that $a\neq 0$) as $n\rightarrow\infty$, for all the initial conditions. The effect of the Kerr nonlinearity is taken into account with a term $-K_S.S_n$ [19], where $K_S$ is here approximated as a constant that includes the pulse intensity and the nonlinear index of refraction. The simplified map describing the spatial effects of the apertures and SF is then:

$$\psi_{n+1} = A - (A+\psi_n)^{-1} - ia - K_S.S_n \qquad (11)$$

Recall that $\psi_n$ is a complex variable: $\psi_n = B\rho_n - 2iBS_n/k$. In the complete (and more accurate) five-variables map, $K_S$ is not a constant and the value of the perturbation is hence much more difficult to calculate.

The study of the values taken by the Lyapunov exponents of (11) shows that, if $A>0$, the map converges to the physically meaningful fixed point [20]. Its basin of attraction covers almost the whole complex plane if $a=0$, and it covers the whole plane if $a>0$. If $A<0$ instead, at least one of the Lyapunov exponents is positive and the iterations diverge. Therefore, if $A<0$ and $(a,K_S)>0$, there are two tendencies in opposition: the effect of the aperture is to converge to the fixed point, while the effect of the nonlinear term is to diverge. Depending on the relative values of $(a,K_S)$ and the initial condition, the map converges to the fixed point, diverges, or converges to periodic orbits. The latter occurs if $(a,K_S.S_n)\ll 1$, i.e., if the last two terms in (11) are small enough to be taken as perturbations of the cold cavity. Most noteworthy, these perturbed periodic orbits accumulate near the orbits of low periodicity of the cold cavity. In other words: the periodic orbits that arise from the balance between the opposite tendencies of the Kerr effect and the aperture, have a period $n$ close to the $P_3^n$ of lowest order having a zero near the cavity's value of $A$ (i.e., this occurs regardless whether $A$ is an exact zero of a low order $P_3^n$ or not). In the Fig.8a, each dot represents the values of $(a, A, K_S)$ of a stable (perturbed) periodic orbit. As it is seen, the set of dots accumulate in vertical lines or "fringes" close to the values of $A$ that are zeroes of polynomials $P_3^n$ of low order. The effect is seen more clearly in the Fig.8b, where only the orbits of period 7 are plotted. The orbits accumulate near the zeroes of $P_3^7$, and slowly shift as $(a,K_S)$ increase.

This effect is quite intuitive, after all. The periodic orbits of the cold cavity survive, slightly modified, to the presence of the two opposite perturbations. The low periodicity orbits dominate, for their iterations are more distant in the complex plane than the ones with higher periodicity, and are hence more robust against the "blurring" of the trajectories caused by the perturbations. As it may be expected, the set of initial conditions leading to periodic orbits have a fractal structure [20].

*4.3 The numbers of the real laser.*

In our laser, the Kerr effect is a small perturbation and there are no tightly closed irises into the cavity. The most significant transversal aperture is the one due to the gain region in the laser rod, which is also a small perturbation once the mode-locking regime is reached. Therefore, the condition that the last two terms in (11) are perturbations of the cold cavity map is fulfilled. The **ABCD** matrix values of the cold cavity can be calculated from the data in Fig.1 and are: $A_0 = 4.1381$, $B_0 = -2.3048$ cm, $C_0 = 8.3276$ cm$^{-1}$, $D_0 = -4.3965$. Therefore: $A = (A_0 + D_0)/2 = -0.1292$. Hence, the condition $A<0$, $K_S>0$ that defines opposite tendencies for the aperture and the Kerr effect is fulfilled too. Finally, the lowest $P_3^n$-polynomial having a zero close to this value of $A$ is $P_3^{11}$ (the zero is -0.142, see the Appendix, the next closest zero of low order is -0.174, of $P_3^9$). It is then reasonable to expect that the "hot" cavity will display orbits with a periodicity near 11, which is precisely what is observed in the lab and also obtained from running the five-variables map.

However, the task is not completed yet. The fact that the actual value of $A$ is close to a zero of $P_3^{11}$ explains why $S_n$, $\rho_n$ follow quasi-periodical orbits of period near 11, but explaining the magic numbers for the ORW requires one further step: the iterations of the cold cavity lie on a circle in the $\{S_n, \rho_n\}$ plane, which is defined by the initial condition. This circle passes between the fixed point and the origin [20]. The iterations are not equally spaced on this circle, and there is always at least one of them in the region between the fixed point and the origin, that is, a region of small $S_n$. From eq.(3e), $U_n S_n \approx$ constant, thus in that iteration a large value of $U_n$ is reached. This process of period 11 is perturbed with the opposite contributions of the aperture and the SF (recall that in the more accurate five-variables map $K_S$ is not a constant), but it survives in the form of the magic numbers, which all are simple combinations of 11 and 12.

To test this explanation, we modify the values of the geometrical parameters of the cavity, to check if the magic numbers change accordingly. We choose $A=-0.225$, which is a root of $P_3^7$ and $P_3^5$. This leads to new values of the cold cavity parameters: $A_0 = 4.1860$, $B_0 = -2.3048$ cm, $C_0 = 8.5058$ cm$^{-1}$, $D_0 = -4.4444$. These values are inserted in the five-variables P2-map and new numerical simulations are run, all the other laser parameters' values being the same as before. In the Fig.9 we display the histogram analogous to Fig.3 for a run of 15000 pulses, with a total of 239 ORW according the 2×AI criterion. As it is seen, the magic numbers are now {5, 7, 12, 17, 22, 23, 24, 26, 27, 121 (only one, not plotted)}. Leaving aside 23 (only one ORW), all these new magic numbers are simple combinations of the numbers 5 and 7, which are the periodicities of the orbits of the cold cavity for the value of $A$ chosen. As it happened before, there are also missing magic numbers: here they are 19 and 29. If the threshold for ORW is lowered, the larger magic numbers disappear and the number of ORW separated by simpler combinations of 5 and 7 increases, in the same way as it was observed in the older cavity. Analogous results are obtained for other values of *A* tested close to other zeroes of the $P_3^n$ of low order.

We conclude then that the magic numbers are the residuals of the stable orbits of the cold cavity, when perturbed by the opposite tendencies of a transversal aperture and SF.

## 5. Summary.

The chaotic regimes of the bistable KLM Ti:Sapphire laser display a large variety of interesting dynamical behaviors, whose exploration is far from being completed here. We have focused here in two intriguing features: *i)* why the ORW appear in the chaotic regime of only one of the two pulsed modes of operation (the chirped-pulse mode or P2), and *ii)* the cause of the discontinuous or quantified distribution of the separation between two successive ORW, as measured in the number of round trips (the "magic numbers").

Regarding *(i)*, evidence is found that the ORW appear after crossing the modulational instability threshold. This occurs first for the mode P2 than for the P1, because the average SPM nonlinearity is larger for the former. Increasing the small signal gain makes the P1 mode to cross the threshold and to display ORW too. Yet, we believe that the reason why ORW are not observed in the mode P1 is that it is unstable at the increased value of the gain. In fact, even if starting at the fixed point value of P1, the system evolves rapidly (in a scale of μs in real time) into the mode P2, where it remains and does display ORW. In other words: we predict that, even if the pump power is increased, no ORW in the P1 mode will be observed in the practice.

Regarding *(ii)*, numerical simulations indicate that the magic numbers are the residuals of the periodic orbits of the cold cavity, when perturbed by the opposite tendencies of an expansive Kerr nonlinearity and dissipative losses. This result was predicted by a simplified theoretical approach that reduces the KLM evolution to an iterative map in the complex plane. The fact that ORW in this laser are generally unpredictable, but that occur at only preferred times, is a further confirmation of their deterministic (i.e., not-noise driven) nature.

It is interesting to note that the formation of ORW in this laser seems related to the modulational instability, that is, it is linked to the *time* part (SPM) of the Kerr nonlinearity. The magic numbers, instead, are related to the distortion of the geometric orbits of the cold cavity due to the presence of apertures and SF, that is, they are linked to the *spatial* part of the Kerr nonlinearity.

The evidence that ORW follow a modulational instability places the extreme events phenomena observed in this laser closer to the rogue waves observed in the ocean. This analogy allows regarding this laser as a promising tool to study general features of the extreme events of this kind, with several practical advantages. Finally, in this paper, we have studied the KLM Ti:Sapphire laser as an object of dynamical interest in itself. Nevertheless, the knowledge obtained from this study may prove helpful to improve the practical performance of this most used device.

**ACKNOWLEDGMENTS.**

Many thanks to Prof. Mario Marconi for his hospitality and helpful assistance during the observations with the Ti:Sapphire laser facility at the Laboratorio de Electrónica Cuántica (LEC), Facultad de Ciencias Exactas y Naturales, Universidad de Buenos Aires. This research was supported by the CONICET (Argentina) contract PIP 2011-077 "Desarrollo de láseres sólidos bombeados por diodos y de algunas de sus aplicaciones", and the AFOSR (USA) grant FA9550-13-1-0120, "Nonlinear dynamics of self-pulsing all-solid-state lasers".

**APPENDIX.**

The first $P_3^n$-polynomials are:

$$P_3^1 = 1$$
$$P_3^2 = 2A$$
$$P_3^3 = 4A^2 - 1$$
$$P_3^4 = 8A^3 - 4A$$

The recursive relationship (10) allows the calculation of the following ones. The zeroes of the $P_3^n$-polynomials until order $n=13$ are:

$n = 1$: no zeroes.

$n = 2$: 0.

$n = 3$: ± ½.

$n = 4$: 0, ± 1/√2.

$n = 5$: ± 0.223, ± 0.901

$n = 6$: 0, ± ½, ± √3/2.

$n = 7$: ± 0.223, ± 0.624, ± 0.901.

$n = 8$: 0, ± 0.329, ± √2/2, ± 0.924.

$n = 9$: ± 0.174, ± ½, ± 0.766, ± 0.940.

$n = 10$: 0, ±0.309, ± 0.588, ± 0.809, ± 0.951.

$n = 11$: ± 0.142, ± 0.415, ± 0.654, ± 0.840, ± 0.960.

$n = 12$: 0, ± 0.259, ± ½, ± √2/2, ± √3/2, ± 0.966.

$n = 13$: ± 0.120, ± 0.350, ± 0.566, ± 0.748, ± 0.883, ± 0.971.

For our laser $A$= -0.129, close to the zero -0.142 of $P_3^{11}$. This is an apparent explanation of the fact that the observed distance between ORW (the "magic numbers") are simple combinations of 11 and 12. This explanation is confirmed by adjusting the cavity parameters to $A$= -0.223, which is a zero of $P_3^5$ and $P_3^7$, what changes the magic numbers to simple combinations of 5 and 7.

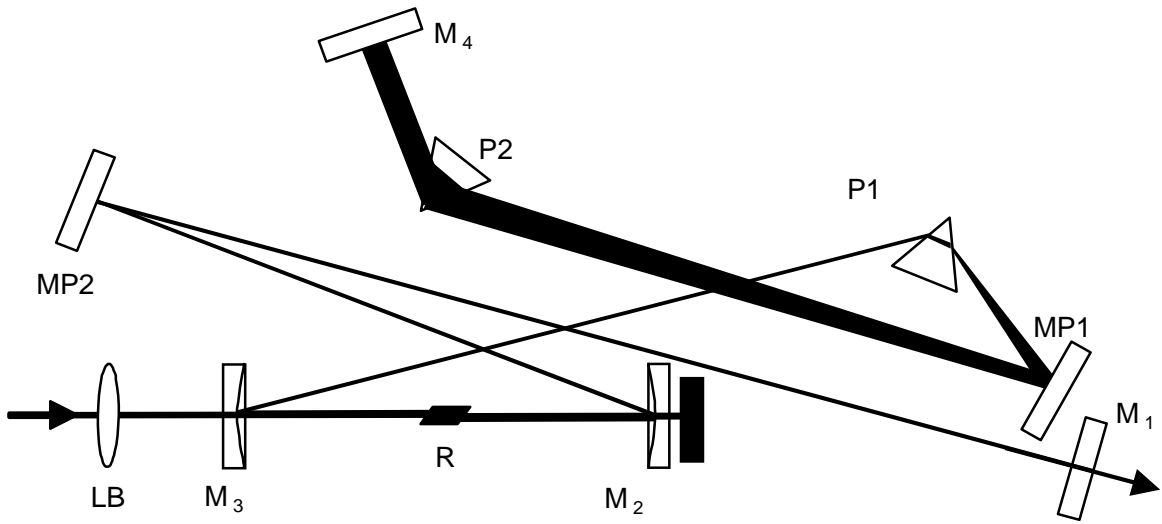

**Figure 1:** Scheme of the laser. LB: pump focusing lens (*f*=10cm), $M_{2,3}$ curved mirrors (R= 10 cm) $MP_{1,2}$ plane HR mirrors, $P_{1,2}$ pair of fused silica prisms to introduce negative GVD. Distances in mm: $M_3$-R= R-$M_2$= 50, $M_2$-$MP_2$= 140, $MP_2$-$M_1$= 465, $M_3$-$P_1$= 297, $P_1$-$MP_1$= 198, $MP_1$-$P_2$= 415, $P_2$-$M_4$= 109.

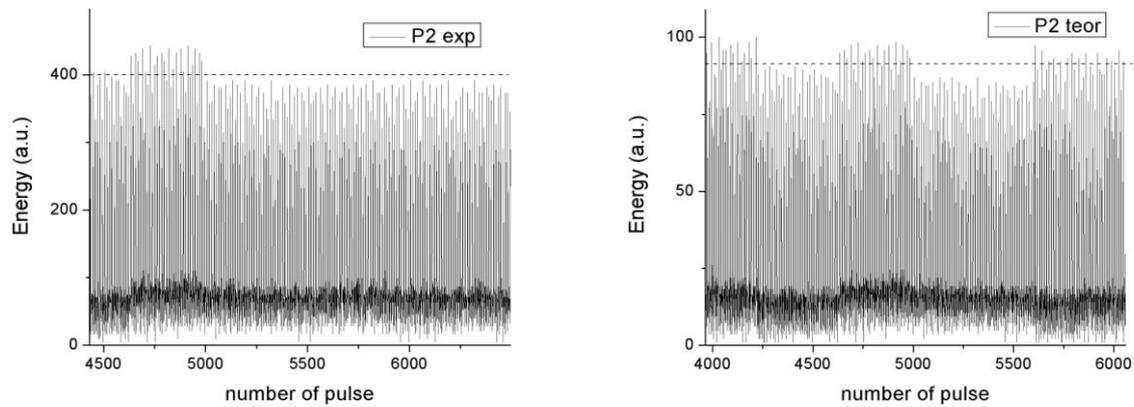

**Figure 2:** ORW in the mode P2. The horizontal dashed line indicates the threshold of ORW for the full time series. The energy scale is in arbitrary units. Left: experimental time series, zoom of ≈2000 pulses of a series of 9978 with a total of 205 ORW, kurtosis= 4.91. Right: Theoretical time series obtained from the five-variables iterative map for the same laser parameters' values, zoom of ≈2000 iterations of a series of $10^4$ with a total of 226 ORW, kurtosis= 4.98. Note the intermittent excursions to a regime of pulses of higher energy in both series.

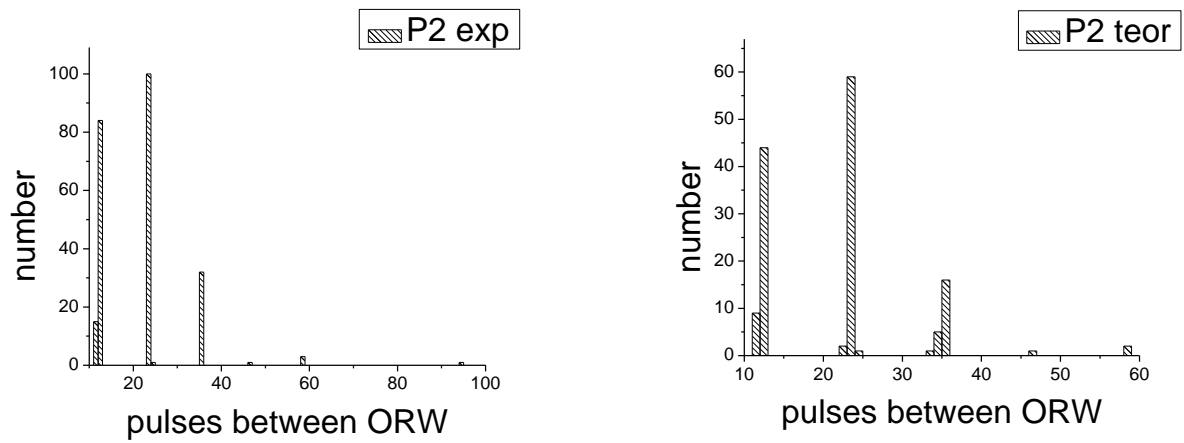

**Figure 3:** Histograms of the number of ORW according to the distance (in cavity round trips) to the next ORW in the chaotic regime of P2; left: experimental results; right: theoretical ones. Average pulse duration: 80 fs.

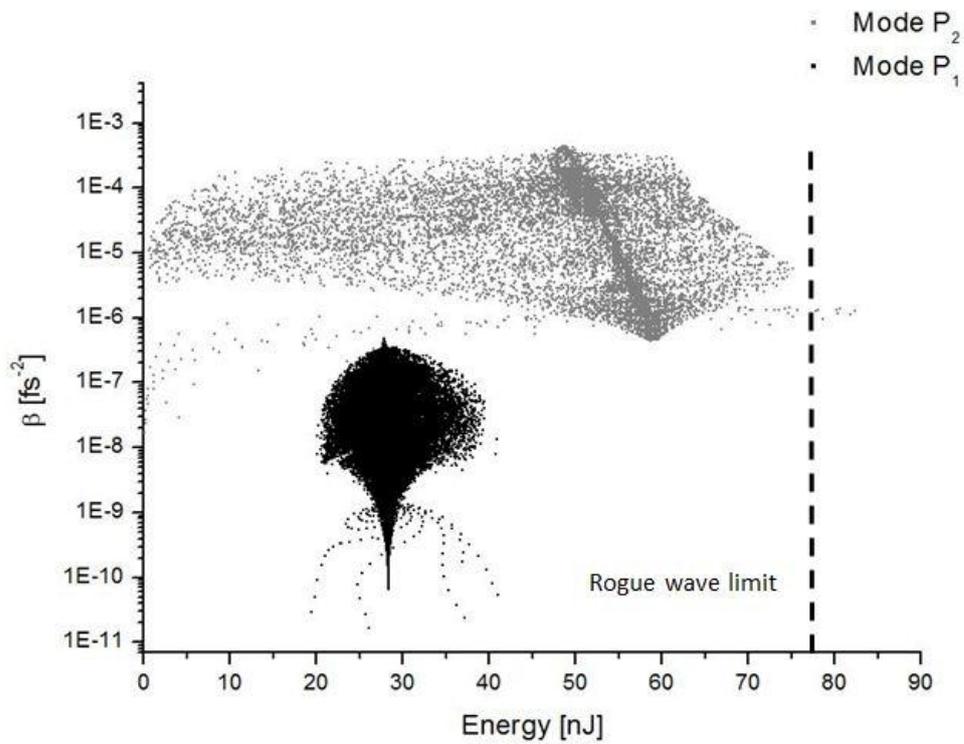

**Figure 4:** Calculated values of the SPM (β) and the energy, each point represents one pulse of a chaotic time series of $10^5$ pulses after discarding a transient of $3\times10^4$. Note that almost all the pulses of the mode P2 (in gray) have a larger value of SPM than the largest ones of the mode P1 (in black). The total number of ORW (the points at the right of the vertical dotted line) is 331, they seem to be fewer because of the scale of the plot. GVD= -42 $fs^2$ for both modes.

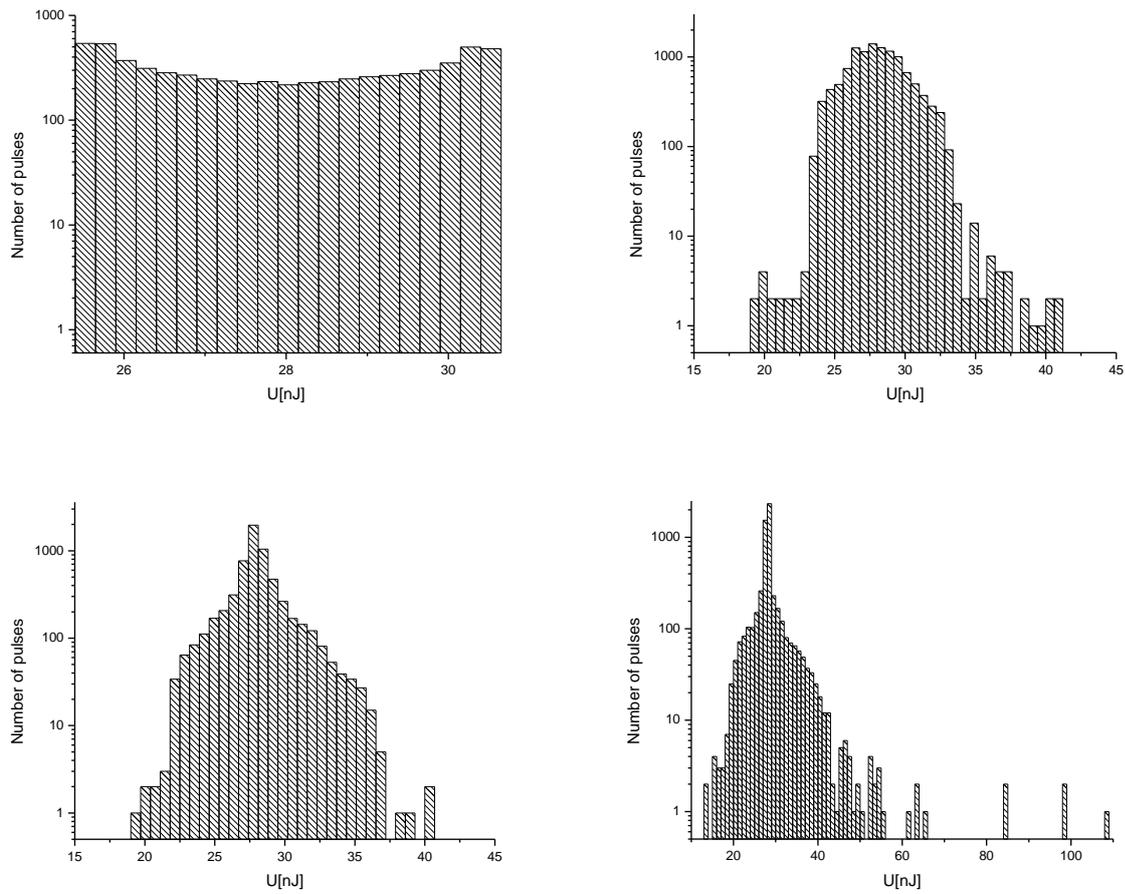

**Figure 5:** Pulse energy histograms (total $10^4$ pulses) in the chaotic regime of the mode P1 for increasing small signal gain, GVD= -42 fs$^2$; (a) operating laser value, (b) 10% increase, (c) 20% increase, (d) 40% increase, a total of 16 ORW are observed, kurtosis= 4.55. For (a-c), the threshold 2×AI>100 nJ, for (d), 2×AI=63 nJ.

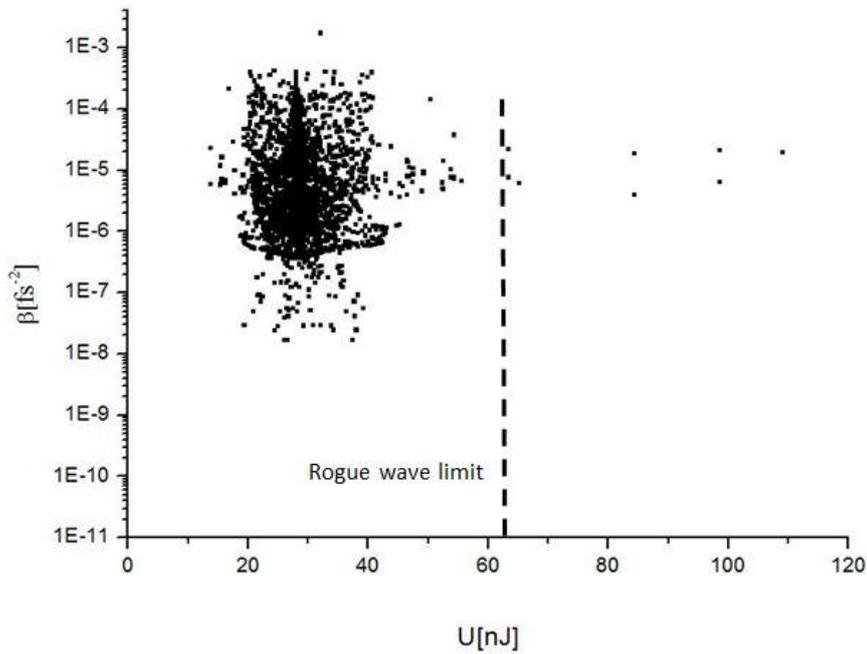

**Figure 6:** Calculated values of the SPM ($\beta$) and the energy as in the Fig.4 but for the P1 mode with a 40% increase in the small signal gain (this plot corresponds to the run in Fig.5d). Note the changes in shape and position of the cloud, and the appearance of ORW. The total number of ORW (the points at the right of the vertical dotted line) is 16, they seem to be fewer because of the scale of the plot. GVD= -42 fs$^2$.

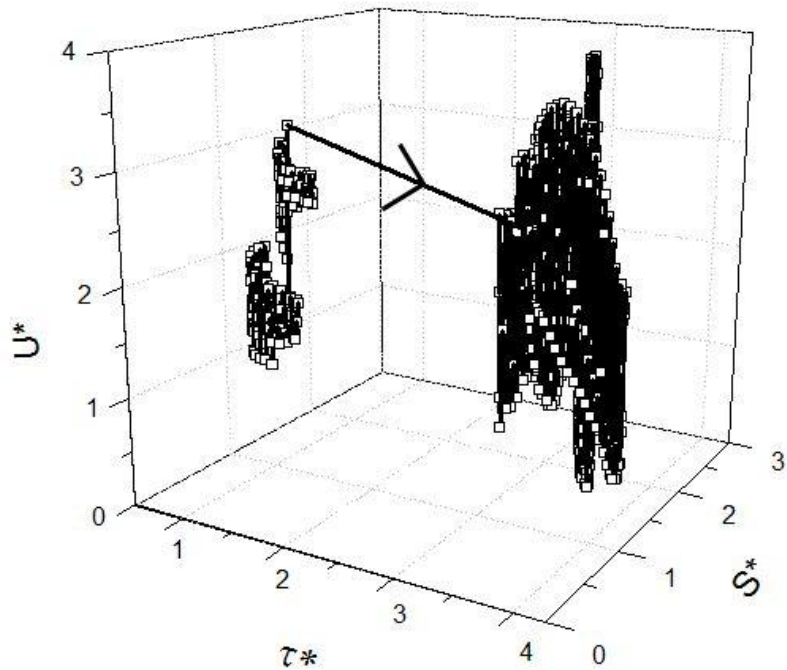

**Figure 7:** Evolution of the pulse starting in the P1 mode (on the left) towards the P2 mode (on the right), according to the "Bistable-map". The parameters correspond to the working point of the real laser, excepting for the small signal gain which is 40% higher. Note that, in this run, a high energy fluctuation of P1 (an ORW?) is immediately followed by a transition to P2, where the system remains (and does display ORW). The initial condition: U=19 nJ, $\sigma$=39 μm, $\tau$=19 fs, Q,$\rho$=0 corresponds to the fixed point of P1, GVD=-42 fs$^2$. The transition to P2 occurs after 218 round trips (approx. 2.5 μs in real time), total length of the run: $10^4$ pulses. The asterisks in the variable names indicate that they are scaled to their values at the fixed point of the mode P1.

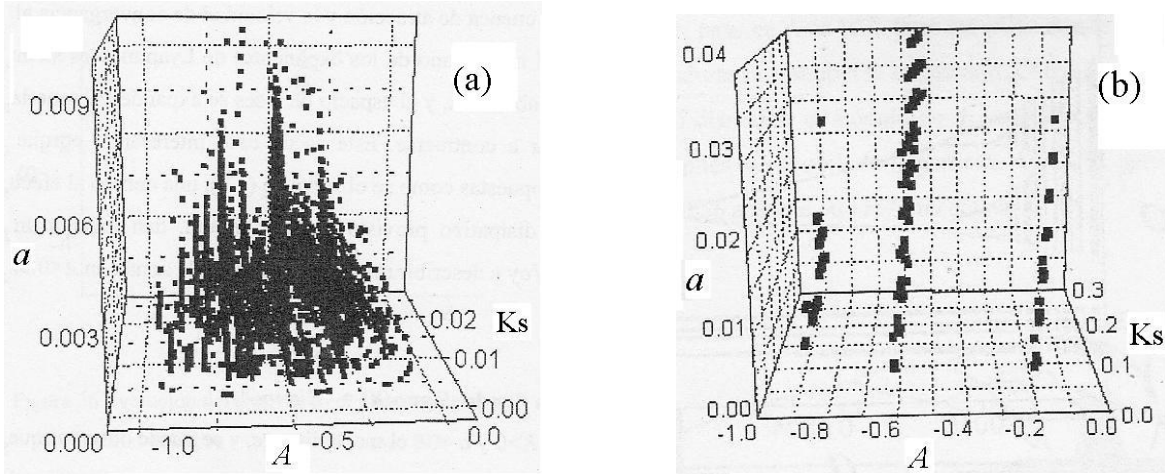

**Figure 8:** (a): values of ($a,A,K_s$) for which stable periodical orbits (period<100) are obtained in the map (11). Each dot represents an orbit, the "fringed" pattern is due to the accumulation of the orbits near the values of $A$ that are zeroes of $P_3^n$ of low order. (b): the same as (a) but orbits of period 7 only, note that they are close to the zeroes of $P_3^7$: {± 0.223, ± 0.624, ± 0.901} for all the values of ($a,A,K_s$) (extracted from Reference 20).

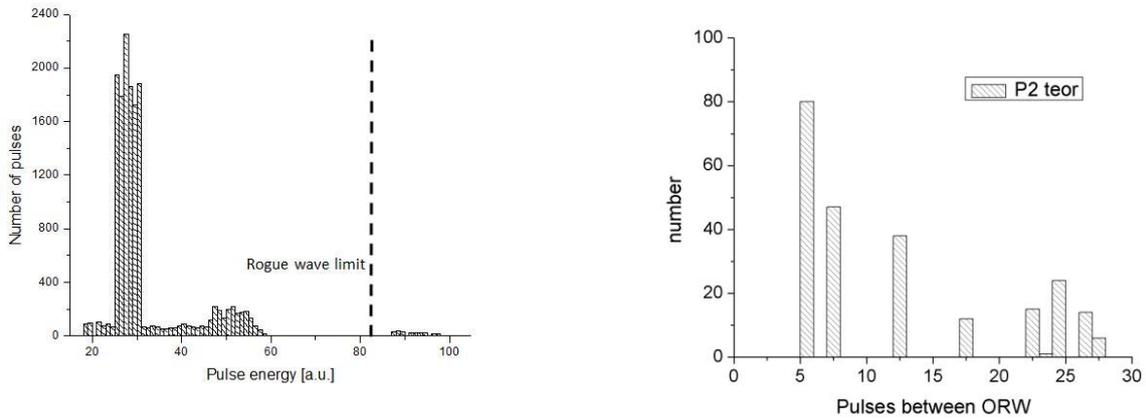

**Figure 9:** Results for the cavity modified such that $A$ is a zero of $P_3^7$ and $P_3^5$. Left: distribution of the number of pulses according to their energy, chaotic regime of the mode P2. The total number of pulses in this run is 15000 and there are 239 ORW. Right: histogram of the number of ORW according to the distance (in cavity round trips) to the next ORW. The magic numbers are now {5, 7, 12, 17, 22, 23, 24, 26, 27, 121 (only one, not shown)}. They are all simple combinations of the numbers 5 and 7, as expected.